 \documentstyle[preprint,prl,aps,epsfig]{revtex}
\def\DESepsf(#1 width #2){\epsfxsize=#2 \epsfbox{#1}}

\def\Bb{\textsf{\textbf{B}}}
\def\Dd{\textsf{\textbf{D}}}

\begin{document}
\pagestyle{empty}                                      
\preprint{
\hbox to \hsize{
\hbox{
            }
\hfill $
\vtop{
 \hbox{NTUHEP-00-25}
 \hbox{BNL-HET-00/30}
 \hbox{ }}$
}
}
\draft
\vfill
%
\title{ Pathways to Rare Baryonic $B$ Decays}
%
\vfill
\author{$^{1}$Wei-Shu Hou and $^2$A. Soni}
\address{
\rm $^1$Department of Physics, National Taiwan University,
Taipei, Taiwan 10764, R.O.C.\\
and\\
\rm $^2$Physics Department, Brookhaven National Laboratory,
Upton, NY 11973, USA}

%
%
\vfill
\maketitle
\begin{abstract}
We point out new ways to search for charmless baryonic $B$ decays:
baryon pair production 
in association with $\eta^\prime$ is very likely
as large as or even a bit larger than two body $K\pi$/$\pi\pi$ modes.
We extend our argument, in weaker form, to
$B\to \gamma + X_s$ and $\ell\nu +X$. 
Although calculations are not reliable,
estimates give branching ratios of order $10^{-5}$--$10^{-6}$, 
where confidence is gained from recent experimental finding that 
$B \to D^{*} p \bar n$, $D^{*} p \bar p \pi$ 
are not far below $D^*\pi$ and $D^*\rho$ rates. 
Observation of charmless baryon modes would help clarify
the dynamics of weak decays to baryonic final states, 
while the self-analyzing prowess of 
the $\Lambda$-baryon can be helpful in $CP$- and $T$-violation studies.

\end{abstract}
\vskip2pc

\pagestyle{plain}

Many charmless mesonic $B \to M_{(s)} \bar M$
decays have emerged at the $10^{-5}$ level in recent years,
giving evidence for strong  $b\to s$ penguins
and tree level $b \to u$ transitions. 
In contrast, the search for rare
baryonic modes has  been less fruitful. 
The most recent limits \cite{baryo} are 
$B\to \bar\Lambda p$, $\bar\Lambda p\pi^-$ and $\bar pp
 < 0.26$, 1.3 and 0.7 $\times 10^{-5}$, respectively,
improving previous bounds \cite{PDG} 
by more than an order of magnitude, 
except for the $p\bar p$ mode which has a  $2.8 \sigma$ excess. 
%
Theoretical work on rare baryonic 
(we use $M$ and $\Bb$ to stand for 
meson and baryon, respectively) 
decay is equally sparse \cite{pole1,pole2,diquark,SR}, 
but in general they predict $B\to \bar\Bb_{(s)}\Bb$ 
to be below $10^{-5}$, oftentimes considerably below.
With the advent of $B$ factories, two body baryonic modes should
eventually emerge. But one may wonder:
%
%
Where is the best place to search for charmless baryonic modes? 

We suggest that charmless baryon-antibaryon final states in $B$ decays 
may show up in association with $\eta'$ and/or $\gamma$ 
with sizeable branching ratios, i.e.\ $\approx 10^{-5} - 10^{-6}$ or more. 
Although theoretical calculations are very unreliable,
the bright side of this is that theory will
learn much from experiment once the measurements become available.
In particular, we would get important input for understanding the
dynamics of weak decays. Furthermore, baryonic final states offer new
observables that should be sensitive probes of $CP$ and $T$ violation.

We take cue from the surprise
discovery of 
the $B\to \eta^\prime + X$ modes.
Without guidance from theory, the CLEO Collaboration 
discovered that both inclusive \cite{etapXs} 
$B\to \eta^\prime + X_s$ (where $X_s = K + n\pi$) and exclusive \cite{etapK}  
$B\to \eta^\prime K$ modes are very large
 ($> 6\times 10^{-4}$ and $\simeq 8\times 10^{-5}$, respectively).
To date they are still not clearly understood. 
Theoretical work done after the fact \cite{ali} can still not fully explain 
the exclusive rate even by some {\it ad hoc} tuning of parameters. 
For the inclusive mode, 
an interesting but still controversial proposal \cite{AS,HT}, 
based on $b\to sg^*$ followed by $g^*\to g\eta^\prime$ transition
(which is motivated by the gluon anomaly),
seems to account for the observed $m_{X_s}$ recoil spectrum and rate.
In the following, we give two semi-quantitative arguments,
one from the inclusive perspective with the anomaly mechanism,
the other from exclusive perspective and based on a pole model,
that suggest $B\to \eta^\prime \Bb_s\Bb(\pi)$~\cite{piparen}
may be comparable to the $\eta^\prime K$ modes with $Br\sim10^{-5}$.
Interestingly,
our suggestion is supported by 
the very recent observation \cite{Cinabro} of 
$B\to D^*p\bar n$ and $D^* p\bar p\pi$ modes
with rates not far below $B\to D^*\pi$, $D^*\rho$.

The partial reconstruction technique used in 
inclusive $B\to\eta^\prime+ X_s$ study was initially developed 
for inclusive $b\to s\gamma$ \cite{bsgamma} studies. 
%
Since both the inclusive $b\to s\gamma \sim 3\times 10^{-4}$
and exclusive $B\to K^*\gamma \simeq 4\times 10^{-5}$~\cite{BK*gamma2} 
are comparable to their $\eta^\prime$ counterparts, 
we extend the pole model argument and suggest that 
$B\to \gamma \bar\Bb_s\Bb(\pi)$ may also be promising. 
Similar arguments suggest that 
$B\to \ell\nu \bar\Bb\Bb(\pi)$ 
should also be searched for.
We also comment on the special case of 
$B\to J/\psi \bar\Bb_s\Bb$ \cite{Brodsky},
where phase space is extremely limited.

Let us understand why $B\to \bar\Bb_{(s)}\Bb(\pi)$ modes 
are so suppressed. 
Baryon formation in $B$ decays is more difficult than
the mesonic case, as reflected in model calculations. 
In pole models \cite{pole1,pole2}, 
the strong $B \to \bar\Bb_{b}\Bb_1$ transition ($\Bb_{b}$ is a $b$-baryon) 
is followed by a $\bar\Bb_{b} \to \bar\Bb_2$ weak transition, 
where the large imbalance in mass is the main source of uncertainty. 
The estimate of transitions to 
final states involving spin 3/2 baryons \cite{pole2} 
seem to be ruled out by experiment already~\cite{baryo}. 
Another intuitive picture 
involves diquarks, which we denote generically as $\Dd$. 
One first has a $b \to \Dd_1 \bar q$ weak decay \cite{diquark}, which, 
together with the spectator quark, gives a $\Dd_1 \bar\Dd_2$ pair. 
Further creation of a $\bar qq$ pairs leads to a $\bar\Bb\Bb$ final state. 
However, since the energy release in $B$ decays is so much larger than 
the argued color $\bar 3$ diquark binding scale $\sim 1$ GeV, 
the approach is dubious for charmless two body final states. 
Finally, the QCD sum rule approach \cite{SR} 
tries to evaluate directly 
the $B$-$\bar\Bb_{(s)}$-$\Bb$ three point function, 
and it is the only method that has studied penguin effects so far. 
The basic problem may be the
applicability of sum rules to $B$ decay to light hadrons. 
Analogous to the diquark picture, 
one relies on a soft $\bar q$-$q$ pair creation model, 
hence it seems to be better suited for 
$B\to \bar\Bb_{c}\Bb$ processes 
where the energy release is lower.

We can now understand why $B\to \bar\Bb_{(s)}\Bb$ modes are 
suppressed compared to $B\to M_{(s)} \bar M$:
Baryons are more complicated objects than
the ``atomic'' mesons and hence harder to form. 
The weak Hamiltonian induces 
$\bar b \to \bar d q\bar q$ and $\bar s q\bar q$ transitions  
that lead to final states already containing two $q\bar q$
pairs, 
typically with matching color that project 
easily onto $M_{(s)} \bar M$ final states.  
In contrast, not all the ingredients are present for 
the $B\to \bar\Bb_{(s)}\Bb$ case. 
The need for creating another $q\bar q$ pair leads to 
suppression by either a strong coupling, 
or by the intrinsic softness of nonperturbative pair creation 
against the rather hard weak decay.

Applying the diquark model to penguin processes 
serves to illustrate further the point. 
The penguin process can be viewed as induced by
a $b\to sg^*$ transition where $g^*$ is virtual. 
One may think that $g^* \to \bar \Dd \Dd$ can be
treated on the same footing as $g^*\to q\bar q$, 
hence $b\to s\bar \Dd \Dd$ could be analogous to $b\to s\bar q q$. 
However, the picture fails because, 
while quarks are truly fundamental, 
diquarks are quark-quark correlations at best up to typical hadronic scale. 
Since the $g^*$ virtuality in two body penguin transitions 
is well above this scale, 
the $g^* \to \bar \Dd \Dd$ transition is suppressed by some form factor. 
Thus, the smallness of  $B\to \bar\Bb_{(s)}\Bb$ modes 
is rooted in the large energy release. 
They are further suppressed compared to $B\to M_{(s)} \bar M$
modes because of the more complicated composition of baryons.

From discussions above, we have gained some insight into where 
charmless baryonic $B$ decays may be larger:
{\it One has to reduce the energy release and at the same time 
allow for baryonic ingredients to be present in the final state}. 
A natural starting point is 
the inclusive $B\to \eta^\prime + X_s$ decay,
where a large rate of $\simeq 6\times 10^{-4}$ is
observed for $p_{\eta^\prime} > 2.0$ GeV\null. 
Much energy is already carried away by the $\eta^\prime$ while 
the signal is established by requiring a cut \cite{etapXs} on 
recoil system mass $m_{X_s} < 2.35$~GeV\null. 
The observed inclusive $m_{X_s}$ spectrum is 
so far accounted for only by the anomaly mechanism \cite{AS,HT}, 
namely $b\to sg^*$ followed by $g^*\to g\eta^\prime$ 
with effective coupling motivated by the gluon anomaly. 
It has been argued that the anomaly coupling should be 
form factor suppressed \cite{AP}
since the $g^*$ is rather virtual ($\sqrt{q^2} \sim 3$~GeV).
However, the problem is interestingly nontrivial \cite{HT}
because of high gluon binding scale 
in the $G_{\mu\nu}\tilde G_{\mu\nu}$ channel,
which has  no analog in the $\gamma^*\to \gamma\pi$ case. 
At any rate, we take this as a model that is rather
effective in producing fast $\eta^\prime$ in $B$ decays. 
We therefore consider the transition 
$\bar b q \to \eta^\prime + \bar s g q$ where 
$\bar s g q$ forms a color singlet.

Treating the gluon as a parton in the final state, 
the $\bar s g q$ system gives an $m_{X_s}$ recoil mass spectrum that is 
in good agreement with data \cite{etapXs} 
and peaks roughly at 2.3 GeV\null.
Although $\bar s g q$ clearly can evolve into $K+ n\pi$, 
it is instructive to visualize how it may feed a single kaon. 
In Ref.~\cite{HT} an effective $m_g = 0.5$ GeV was used 
in final state phase space to remove ``soft'' gluons 
(below consituent $q\bar q$ threshold). 
Since there are no infrared singularities, 
it was pointed out that the $m_{X_s}$ region
covered by $m_g \lesssim 0.5$ GeV 
might be swept under the kaon, and could 
by itself account for the observed size of $B\to\eta^\prime K$. 
Such ``$\vert \bar s g q\rangle$ Fock component of the $K$ meson" 
contributions to $B\to \eta^\prime K$
have not been taken into account in the traditional 
approach \cite{ali}. 
Here we extend this observation and 
exploit $g^* \to \bar \Dd \Dd$ diquark pair creation
to construct baryonic final states.

We illustrate the $\bar b q \to \eta^\prime + \bar s g q$ transition
and $\bar s g q \to \bar s \bar \Dd \Dd q$ evolution
in Fig.~1(a) and (b), respectively. 
Although the $g$-$\bar \Dd$-$\Dd$ coupling can be quantified, 
we shall refrain from introducing further model dependence in 
baryon formation from, say, $\Dd q$. 
Rather, we shall utilize mainly kinematic arguments.
Following the example of how a single $K$ may emerge from $\bar s g q$, 
we take $m_g$ as an effective mass in final state phase space to
correspond to $\bar qq$ or $\bar \Dd\Dd$ formation thresholds. 
For $m_g \lesssim 0.6$ GeV $\lesssim 2m_q$,
where $m_q$ is the constituent quark mass, 
one can view that the gluon ``has no place to go'',  
hence it could only end up in the $K$ meson, as argued earlier. 
For 0.6 GeV $\lesssim m_g \lesssim$ 1.1~GeV $\lesssim 2m_s \sim 2m_\Dd$ 
 (we treat $m_s$ and $m_\Dd$ as roughly equal), 
the ``gluon'' can only split into $u\bar u$ and $d\bar d$,
and one has non-resonant formation of $K\pi$, $K2\pi$, etc.,
or the formation of $\vert K_g\rangle = \vert \bar s g q\rangle$
hybrid mesons.
For $m_g \gtrsim$ 1.1 GeV, 
not only $u\bar u$, $d\bar d$ and $s\bar s$ are open,
so is $\bar \Dd \Dd$. 
Until effective $m_g$ becomes very massive, say beyond 1.8~GeV, 
diquark pair formation is on similar footing with $q\bar q$ 
and is not form factor suppressed.


We depict in Fig.~2 the regions separated by 
$m_g = 0.6$, 1.1 and 1.8 GeV\null.  
Counting spin degrees of freedom only, 
we estimate that roughly 1/13 and up to to 3/7
(depending on scalar or vector diquark nature) of the rate in 
the 1.1~GeV $\lesssim m_g \lesssim$ 1.8 GeV domain 
corresponds to $\bar s \bar \Dd \Dd q$ final state. 
For $m_{X_s} < $ 2.3~GeV (the recoil peak in anomaly model),
phase space and kinematics considerations suggest that 
this final state would preferentially end up in 
two body $\bar\Bb_{s}\Bb$ final states,
given that the $\bar\Lambda p$ threshold is at 2.05 GeV. 
It is likely that one would receive threshold enhancement
since diquark pairs are already produced to the right of the
$m_g \simeq$ 1.1 GeV curve.
The modes to search for are therefore $B\to \eta^\prime \bar\Lambda N$ 
and similar low lying $\bar\Bb_{s}\Bb$ states
accompanying a relatively fast $\eta^\prime$.
We stress that the reconstruction of $\eta^\prime\bar\Bb_{s}\Bb$ modes 
should be easy and with little background,
since the $\Lambda_c^+\bar N$ threshold does not open up until 3.22 GeV.  
It therefore may offer an important probe into 
the higher mass $m_{X_s}$ spectrum
not afforded by $X_s = K+n\pi$ modes.
One can in principle pursue the inclusive study of 
$B\to \eta^\prime \bar\Lambda N + n\pi(\gamma)$,
where $\gamma$ could come from e.g. $\Sigma\to \Lambda\gamma$.

The picture we outline above bears some similarity with
the explanation for the low $p_{J/\psi}$ ``bump" 
in the feed-down subtracted inclusive primary $J/\psi$ momentum 
spectrum seen by CLEO \cite{charmonium}.
The excess for $p_{J/\psi} < 0.6$~GeV is suggested \cite{Brodsky}
to be $B\to J/\psi \bar\Lambda p$ where there is 
only 128 MeV available kinetic energy.
It is known that $B\to J/\psi +X$ decay has a large component 
coming from $\bar cc$ color octet.
Although this excess color is not necessarily shed by a single gluon
as in our anomaly model diagram,
kinematic arguments were also used to argue for 
$\bar\Lambda p$ in final state.
The enhancement may come about because 
nonperturbative effects are operative for such low kinetic energy.
It should be noted that, because of the latter, 
the detection of $\bar\Lambda p$ system recoiling
against $J/\psi$ would not be easy.
In the anomaly model mechanism for explaining 
fast $\eta^\prime$ production in $B$ decays as we outlined above,
the $X_s$ recoil system has $m_{X_s}$ peaked at 2.3 GeV.
On the one hand this is not far above $\bar\Lambda p$ threshold
so one again does not expect the opening of many channels.
On the other hand, the $\bar\Lambda$ and $p$ baryons have 
considerable kinetic energy since they are recoiling against
an energetic $\eta^\prime$, the energy of which is 
(conjectured to be) fed by the $g^*g\eta^\prime$ vertex.
Thus, discovery of $B \to \eta^\prime\bar\Bb_{s}\Bb$ modes
with energetic $\eta^\prime$ may be more straightforward 
than detecting $B\to J/\psi \bar\Lambda p$,
and it may also give credence to the anomaly mechanism itself.


An alternative approach offers complementary support for 
our discussion above from a different perspective. 
Using naive factorization and simple pole model ideas, 
the $B \rightarrow \eta^\prime \bar\Lambda p$ decay is seen (Fig. 3(a))
as occurring in two steps: 
$B \rightarrow \eta^\prime + $``$K$'', 
followed by ``$K$''$ \rightarrow \bar \Lambda p$, 
where ``$K$'' denotes an off-shell kaon. 
The first vertex can be normalized to 
the observed rate for $B \to \eta^\prime K$,
but it is very difficult to make reliable statements about 
the strength of the dimensionless $K$-$p$-$\Lambda$ effective
coupling, $g_{beff}$. 
The crude approximation of $g_{beff}^2/4 \pi \approx 0.3$ gives
${\Gamma(B \to \eta^\prime\bar\Lambda p)/\Gamma(B \to \eta^\prime K)} 
\approx 0.3$,
comparable to the estimate made above through diquark arguments.

Although the simple factorization and pole model 
ideas are far from reliable,
fortunately, some very recent results from CLEO
offer support for the above number.
Following a suggestion by Dunietz \cite{Dunietz},
the first exclusive $B$ decays to nucleons have just been observed,
with~\cite{Cinabro} 
$B^0\to D^{*-} p \bar n$, $D^{*-} p\bar p\pi^+ 
\approx 1.45 \times 10^{-3}$, $6.6 \times 10^{-4}$, respectively. 
As illustrated in Fig. 3(b),
starting from $B^0 \to D^{*-} \pi^+,\ D^{*-} \rho^+ \approx 
2.8 \times 10^{-3}$, $6.7 \times 10^{-3}$,
with effective couplings analogous to $K$-$p$-$\Lambda$ case above,
one easily attains order of magnitude understanding 
of the large strength of $D^{*-} N \bar N(\pi)$ modes.
In general, CLEO's observation of 
sizable exclusive $B\to \bar D^{*} N \bar N(\pi)$ decays
that are not far below $B \to \bar D^{*} \pi$
gives strong support to our argument
that $B \to \eta^\prime\bar\Lambda N(\pi)$
may not be far lower than $B \to \eta^\prime K$!


The situation with regard to $B \to \bar \Lambda p \gamma$ 
is quite similar in the pole model picture,
except $K$ is replaced by $K^*$.
Similar estimates as above again give 
$\Gamma(B\to \bar\Lambda p\gamma)/\Gamma(B\to K^\ast\gamma) 
\approx 0.1$--$0.3$.
These estimates place 
the baryonic branching ratios with $\eta^\prime$ and $\gamma$ 
in the range of $10^{-5}$--$10^{-6}$ and therefore within
reach of the luminosities of the $B$-factories. 
As in the $\eta^\prime \bar\Lambda p$ case,
we stress that the final states $\gamma \bar\Lambda p (\pi)$, 
with energetic photon characteristic of $b \to s \gamma$, 
are reconstructible, clean and should have little background.

It is tempting to extend the pole model picture further into
$B\to \ell^+\nu + \bar \Bb\Bb(\pi)$ via $\pi/\rho$ poles.
In principle, similar numbers would follow.
We caution, however, that, 
unless $E_\ell$ or $m_{\ell\nu}$ are large,
the energy release may not be reduced sufficiently.
Furthermore, charmed baryon production in semileptonic $B$ decays,
where one has reduced energy release,
is relatively suppressed \cite{semibaryon}.
In particular, $\Gamma(B\to\bar pe^+\nu +X)/
                \Gamma(B\to\ e^+\nu +X) < 0.015$ \cite{semibaryon}
suggest that $B\to \ell^+\nu + \bar \Bb\Bb(\pi)$
may be less promising than in association with 
fast $\eta^\prime$ or $\gamma$.

Before we conclude, let us review, in descending order of inclusive rate,
the processes to be studied for charmless baryonic $B$ decays.
The $B\to J/\psi + X$ process is at 1\% level. 
In some sense it is not really a charmless final state,
and has very limited phase space for $X = \bar\Lambda p$ 
(the {\it only} possibility).
Interestingly, a distortion or a bump in the low $J/\psi$ momentum region
is found, indicating that $B\to J/\psi\bar\Lambda p$ could be of order
4\% -- 5\% of $B\to J/\psi + X$ rate.
But this mode is not easy to reconstruct because of the very slow proton.
The charmless $B\to \ell^+\nu + X_u$ process is at $10^{-3}$ level,
but, as mentioned, a bound already exists indicating that
charmed baryon content of $B\to \ell^+\nu + X_c$ is less than 1.5\%
of semileptonic rate.
Hence charmless baryon content of 
semileptonic decays may be less promising.
The $B\to\eta^\prime + X_s$ process is at $10^{-3}$ level,
and, if anomaly mechanism is correct, we expect
$X_s$ to be of order 10\% or more composed of
$\bar \Bb_s\Bb(\pi)$,
with supporting arguments from pole models
as well as the newly observed $B\to D^* N\bar N(\pi)$ modes.
Hence $B\to \eta^\prime \bar \Bb_s\Bb(\pi)$ should be searched for.
The $B\to \gamma + X_s$ is at $3\times 10^{-4}$ level.
Although one only has the pole model argument to suggest the possibility of
sizable baryonic recoil, it again should be studied.

It should be stressed that,
besides yielding useful information for 
theory about the dynamics of weak decays, 
the observation of baryonic final states suggested here
could open up a new direction of studies.
It is well known that the decay of the $\Lambda$ self-analyzes its spin, 
which can be important for studying $CP$- or $T$-violation.
In particular, use of the $\Lambda$ spin will allow one to construct
$CP$-odd, $T_N$-odd observables 
such as 
{\boldmath $s$}$_{\bar \Lambda}\cdot$
({\boldmath $p$}$_p \times${\boldmath $p$}$_{\bar \Lambda}$),
which are
driven by the real (or dispersive) part of the Feynman amplitude 
and not by the imaginary (or absorptive) part of the amplitude, 
as is the case for a $CP$-odd, $T_N$-even observable, 
such as partial rate asymmetries.
Since $\eta^\prime$ and photonic modes are dominantly loop-induced 
hence sensitive windows on new physics,
such probes may turn out to be very useful.

In conclusion,
we point out that $B\to\eta^\prime + \bar \Bb_s\Bb(\pi)$
could be the most promising charmless baryonic modes.
The $\eta^\prime$ should still be fast,
and for $m_{X_s}$ that is not far above $\bar\Lambda p$ threshold,
the baryonic recoil system is simple and of low multiplicity. 
These modes not only could be the
first charmless baryonic modes to be detected,
their detection could strengthen the anomaly picture, 
and provide new probes for $CP$ and $T$ violation
by bringing in the powerful self-analyzed $\Lambda$ spin observable.
Though the argument gets a bit less compelling,
a parallel program should also be started to reconstruct 
baryonic modes in the recoil system against the photon in
$B\to \gamma + X_s$.
The traditional search for two body $B\to \bar \Bb_{(s)}\Bb$ modes
of course should continue,
but observation of rare charmless baryonic $B$ decays proposed here
could open a new program for the study of 
interplay of weak and strong dynamics,
and offer very important probes of new $CP$ and $T$ violation observables.

This work is supported in part by
grants NSC 89-2112-M-002-036 of the Republic of China
and the US DOE contract number DE-AC02-98CH10886(BNL).
We thank Jim~Alexander, Tom Browder,
Tony Rubiera and John Yelton for discussions.

\begin{figure}[htb]
\centerline{\DESepsf(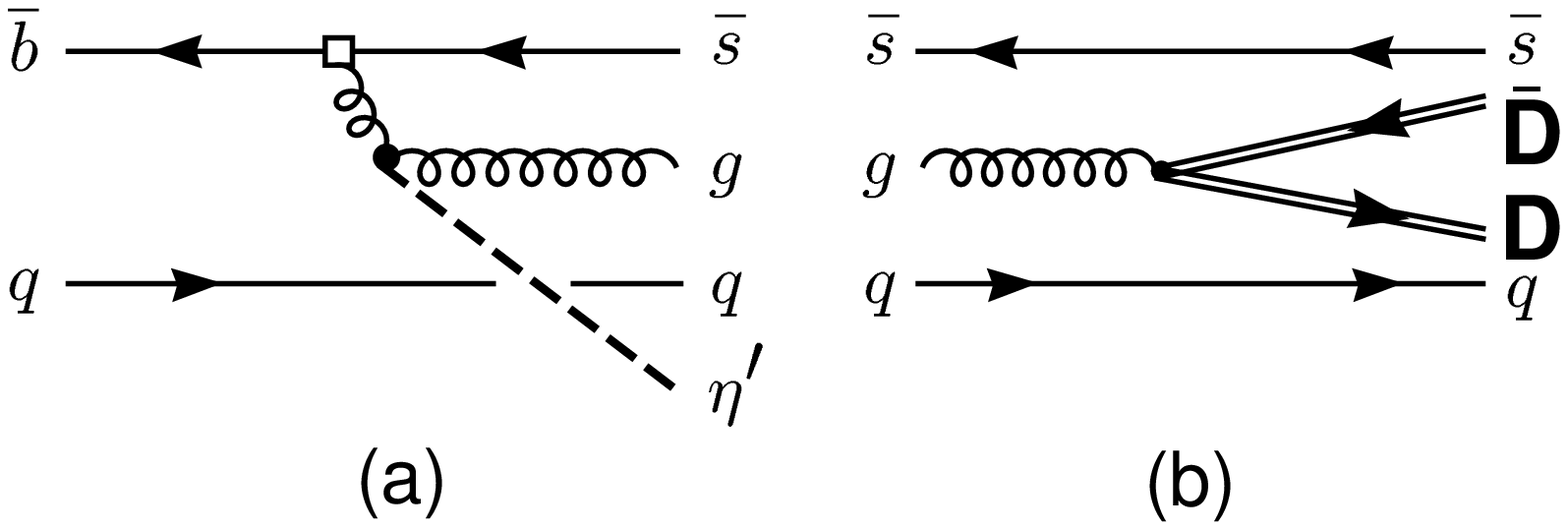 width 10cm)}
\smallskip
\caption{Diagrams illustrating baryon formation via
(a) the $\bar b q \to \eta^\prime + \bar s g q$ transition,
followed by (b) $\bar s g q \to \bar s \bar \Dd \Dd q$ evolution,
where $\Dd$ denotes a diquark.}
\end{figure}
\vskip-0.5cm

\begin{figure}[htb]
\centerline{\DESepsf(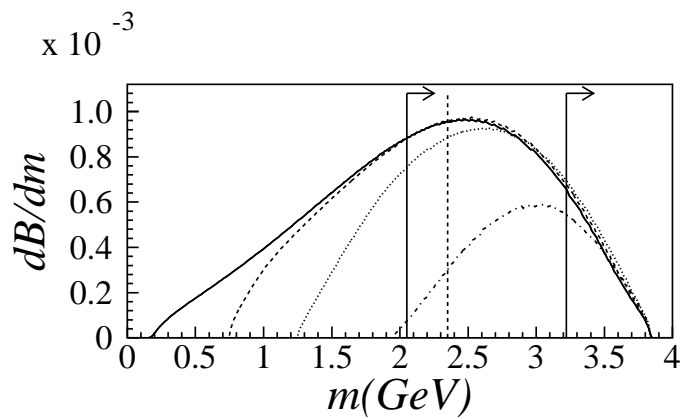 width 10cm)}
\smallskip
\caption{Illustration for phase space argument of 
$\eta^\prime +$ baryon pair formation 
following the mechanism of Fig. 1.
Solid, dash, dots and dotdash curves correspond to
taking $m_g = 0,\ 0.6,\ 1.1,\ 1.8$ GeV in phase space.
The ``gluon mass" of 0.6 (1.1) GeV mark the opening of 
$u\bar u$, $d\bar d$ ($s\bar s$, $\bar \Dd \Dd$) thresholds, 
while beyond $1.8$ GeV, 
diquark pair formation may suffer from form factors. 
The dotted vertical line indicates the experimental 
cut on $m \equiv m_{X_s}$ in $B\to \eta^\prime + K+n\pi$ search,
while the two solid vertical lines to the left and right
corresponds to the $\bar \Lambda N$ and $\Lambda_c^+ \bar N$
thresholds of 2.05 and 3.22~GeV, respectively.
} 
\end{figure}
\vskip-0.5cm

\begin{figure}[htb]
\centerline{\DESepsf(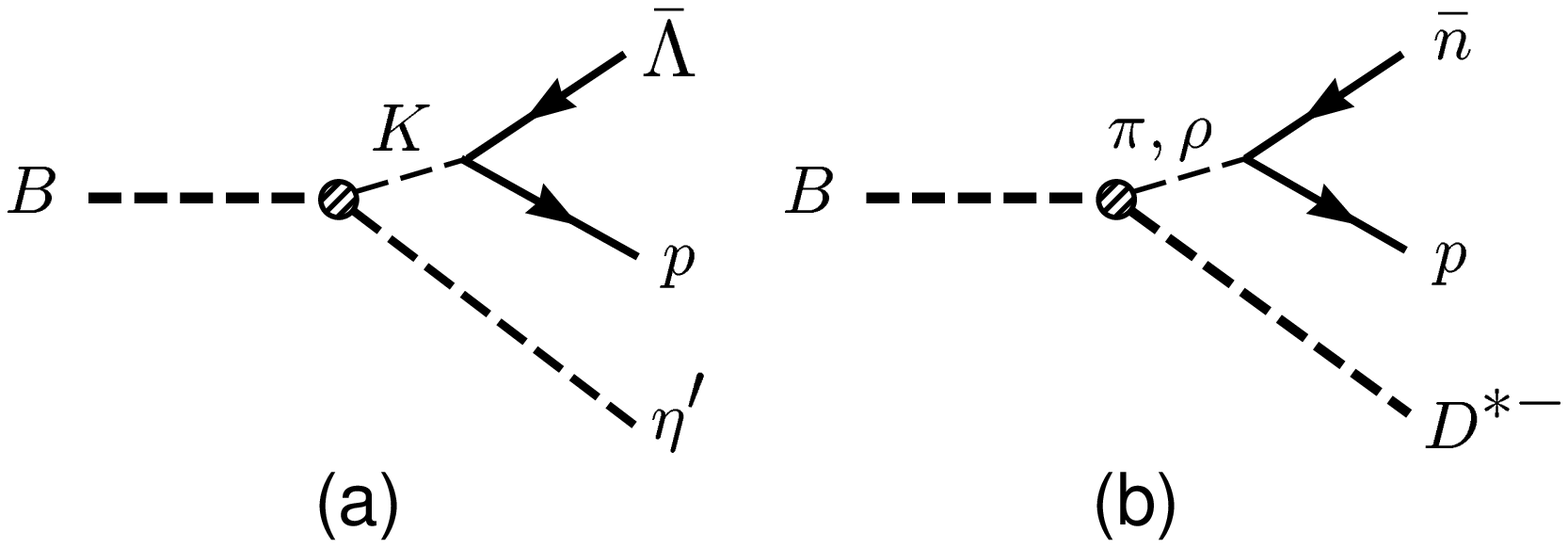 width 10cm)}
\smallskip
\caption{Pole model diagrams illustrating 
(a) $B\to \eta^\prime\bar\Lambda p$ and 
(b) $B\to D^{*-}\bar n p$ production
as mediated by $B\to \eta^\prime K$, and
$B\to D^{*-}\pi$, $D^{*-}\rho$,
respectively.} 
\end{figure}
\vskip-0.5cm


\begin{references}
%
%

\bibitem{baryo} T.E. Coan {\it et al}. (CLEO Collaboration), 
Phys.\ Rev.\ D {\bf 59}, 111101 (1999).

\bibitem{PDG} C. Caso {\it et al}.\ (Particle Data Group), 
Eur.\ Phys. J. C. {\bf 3}, 1 (1998). 

\bibitem{pole1} N. Deshpande, J. Trampetic and A. Soni, 
Mod.\ Phys.\ Lett.\ {\bf 3 A}, 749 (1988). 

\bibitem{pole2} M. Jarfi {\it et al}., 
Phys.\ Lett.\ {\bf B237}, 513 (1990);
Phys.\ Rev.\ D {\bf 43}, 1599 (1991).  

\bibitem{diquark} P. Ball and H.G. Dosch, 
Z. Phys.\ {\bf C51}, 445 (1991). 

\bibitem{SR} V. Chernyak and I. Zhitnitsky, 
Nucl.\ Phys.\ {\bf B345}, 137 (1990). 

\bibitem{etapXs} T.E. Browder {\it et al}. (CLEO Collaboration),
Phys.\ Rev.\ Lett.\  {\bf 81}, 1786 (1998).  

\bibitem{etapK} B. Behrens {\it et al}. (CLEO Collaboration),
Phys.\ Rev.\ Lett.\  {\bf 80}, 371 (1998).   

\bibitem{ali} See, for example, A. Ali, G. Kramer and C.D. L\" u,
Phys.\ Rev.\ D {\bf 58}, 094009 (1998);
Y.H. Chen et al., {\it ibid.} D {\bf 60}, 094014 (1999). 

\bibitem{AS}  D. Atwood and A. Soni, 
Phys.\ Lett.\ {\bf B 405}, 150 (1997). 

\bibitem{HT} W.S. Hou and B. Tseng, 
Phys.\ Rev.\ Lett.\ {\bf 80}, 434 (1998). 

\bibitem{piparen} The $\pi$ in the parenthesis is to serve as a
reminder that the reaction can occur at an exclusive (no $\pi$) and also
at an inclusive ($n\pi$) level.   

\bibitem{Cinabro} D. Cinabro,
plenary talk at ICHEP2000, July 27 -- August 2, 2000, Osaka, Japan;
A.I. Rubiera, U. Florida dissertation, August 2000 (unpublished).

\bibitem{bsgamma} M.S. Alam {\it et al}. (CLEO Collaboration),
Phys.\ Rev.\ Lett.\  {\bf 74}, 2885 (1995).  

%
%

\bibitem{BK*gamma2} T.E. Coan {\it et al}. (CLEO Collaboration), 
Phys.\ Rev.\ Lett.\  {\bf 84}, 5283 (2000).

\bibitem{Brodsky} S.J. Brodsky and F.S. Navarra, 
Phys.\ Lett.\ {\bf B 411}, 152 (1997). 

\bibitem{AP} A.L. Kagan and A.A. Petrov, e-print hep-ph/9707354.  

\bibitem{charmonium} R. Balest {\it et al}. (CLEO Collaboration),
Phys.\ Rev.\  D {\bf 52}, 2661 (1995).  

\bibitem{Dunietz} I. Dunietz,
Phys.\ Rev.\ D {\bf 58}, 094010 (1998);

\bibitem{semibaryon} G. Bonvicini {\it et al}. (CLEO Collaboration),
Phys.\ Rev.\  D {\bf 57}, 6604 (1998).  

\end{references}
\end{document}